# Epistemic Logic for Communication Chains

Jeffrey Kane

Department of Mathematics
and Computer Science
McDaniel College
Westminster, Maryland, USA
jmk001@mcdaniel.edu

Pavel Naumov

Department of Mathematics
and Computer Science
McDaniel College
Westminster, Maryland, USA
pnaumov@mcdaniel.edu

## ABSTRACT

The paper considers epistemic properties of linear communication chains. It describes a sound and complete logical system that, in addition to the standard axioms of $S_5$ in a multi-modal language, contains two non-trivial axioms that capture the linear structure of communication chains.

## 1. INTRODUCTION

In this paper we study epistemic properties of linear communication protocols that we call *communication chains*. An example of such a protocol is the Telephone game[1] depicted in Figure 1: person $P$ picks a random four-letter word $a$ and communicates it to $Q$. Person $Q$ changes at most one letter in $a$, and communicates it to person $R$ as $b$. Finally, $R$ again changes at most one letter in $b$ and communicates it to $S$ as $c$. For instance, sequence $(a,b,c)$ could be $(byte, bite, cite)$. We refer to such a sequence as a *run* of the protocol.

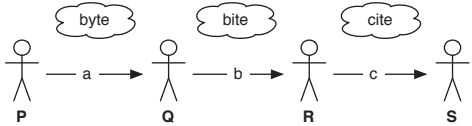

**Figure 1: Telephone Game.**

Note that anyone who knows the value of message $a$ on the run $r_1 = (byte, bite, cite)$ will be able to conclude that $c \neq book$. We say that channel $a$ on run $r_1$ "knows" that $c \neq book$ and write it as $r_1 \Vdash \Box_a(c \neq book)$. Note also that anyone who knows the value of $a$ on the run $r_1$ will also be able to conclude that anyone knowing the value of $b$ on the same run will be able to conclude that $c \neq book$. We write this as $r_1 \Vdash \Box_a \Box_b (c \neq book)$.

Formulas that are true on one run might not be true on another run of the same protocol. For example, if $r_2 = (toon, torn, tort)$, then $r_2 \nVdash \Box_a(c \neq book)$ since a person who only knows the value of $a$ on run $r_2$ cannot distinguish this run from $(toon, boon, book)$. One can consider statements that are true on any run of the Telephone game protocol. Examples of such statements are:

$$\Box_b(a \neq book) \to \Box_b(c \neq book),$$

$$\Box_a(c \neq book) \to \Box_b(c \neq book).$$

---
[1]This game is also known as Chinese Whispers, Grapevine, Broken Telephone, Whisper Down the Lane, and Gossip.

The first of these statements is true due to the symmetry of the Telephone game: if $(a,b,c)$ is a run then $(c,b,a)$ is also a run. This property is not necessarily true for all protocols. The second statement, although it is written in the language specific to the Telephone game, can be stated in the form which is true on each run of each protocol over the communication chain depicted in Figure 1:

$$\Box_a p_c \to \Box_b p_c, \tag{1}$$

where $p_c$ is an arbitrary atomic proposition about the value of the message $c$. In this paper we study that type of "universal" statements that are true on each run of each protocol.

As we will see later, runs can be viewed as Kripke worlds, so all formulas provable in multi-modal version of $S_5$ are "universal" statements in our sense. In addition to $S_5$ theorems, however, our logical system included many facts that reflect the linear structure of the communication chain. The above formula (1) is one of them. Other, less obvious examples are:

$$\Box_a \Diamond_c \varphi \to \Box_b \Diamond_c \varphi,$$

$$\Box_a \Box_c \varphi \to \Box_a \Box_b \Box_c \varphi,$$

$$\Box_b(\Box_a \varphi \vee \Box_c \psi) \to (\Box_b \varphi \vee \Box_b \psi),$$

where $\varphi$ and $\psi$ are arbitrary formulas and $\Diamond_c$, as usual in modal logic, stands for $\neg \Box_c \neg$. We will prove soundness of these principles in Section 4.

The main result of this paper is a sound and complete axiomatization of all properties that are true on each run of each protocol of a given communication chain.

A communication chain can also be interpreted as a timeline. Then, formula $\Box_k \varphi$ means that anyone, who has complete information about a moment $k$ in history, knows that $\varphi$ is true. For example, one can say,

$\Box_{2012}$(In the past, dinosaurs roamed the Earth) $\to$

$\Box_{2011}$(In the past, dinosaurs roamed the Earth).

This interpretation connects our work with other works on axiomatizations reasoning about time [2, 3, 8, 10, 11]. These works, however, are very different from ours in the syntax and semantics that they use. Properties like the the three formulas above cannot be expressed in their language.

Epistemic logic for reasoning about communication graphs, in a language significantly different from ours, was proposed by Pacuit and Parikh [9]. They prove decidability of their logical system, but do not give a complete explicit axiomatization.

This work is also connected to works on information flow on graphs [1, 4, 5, 6, 7], that study properties of nondeducibility, functional dependency, and fault tolerance predicates. Unlike these works, this paper is using modal language. We discuss possible generalization of our work to arbitrary communication graphs in the conclusion.

## 2. SYNTAX AND SEMANTICS

In the informal discussion above, we have implicitly assumed that communication chains have finite length. In the formal presentation through the rest of the paper we consider infinite chains whose communication channels are labeled by consecutive integer numbers (see Figure 2). This is done in order to simplify our presentation. Our results still hold for finite chains. Furthermore, any finite chain can be viewed as an infinite chain in which a fixed default message is sent through a cofinite number of channels.

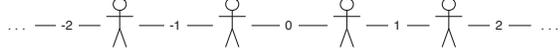

**Figure 2: Infinite Chain.**

We also assume that for each $k \in \mathbb{Z}$ there is a (possibly infinite) set $P_k$ of "atomic propositions" about channel $k$ and that sets $P_k$ and $P_m$ are disjoint for each $k \neq m$.

Next we define formulas in our language. The set of all formulas will be denoted by $\Phi(\mathbb{Z})$. By $\Phi(A)$ we denote the set of formulas whose "outermost" modalities have form $\Box_k$ for some $k \in A$ and "outermost" atomic propositions belong to $P_k$ for some $k \in A$. Thus, for example,

$$\Box_k(\Box_m \varphi \to \Box_n \psi) \in \Phi(\{k\})$$

$$\Box_m \Box_k \varphi \to \Box_n \Box_k \psi \in \Phi(\{m,n\}).$$

DEFINITION 1. *For each $A \subseteq \mathbb{Z}$, set $\Phi(A)$ is the minimal set of formulas such that*

1. $\bot \in \Phi(A)$,
2. $P_k \subseteq \Phi(A)$, *for each* $k \in A$,
3. *if* $\varphi \in \Phi(A)$ *and* $\psi \in \Phi(A)$, *then* $\varphi \to \psi \in \Phi(A)$.
4. *if* $\varphi \in \Phi(\mathbb{Z})$, *then* $\Box_k \varphi \in \Phi(A)$, *for each* $k \in A$.

We assume that the boolean connectives $\wedge$, $\vee$, and $\neg$ are defined through $\to$ and $\bot$ in the standard way. As common in modal logic, by $\Diamond_k \varphi$ we denote formula $\neg \Box_k \neg \varphi$.

In the Telephone game example in the introduction, we have assumed that all messages are four-letter words. In general, we will allow each channel $k$ to have its own set of possible values $V_k$. In the same example, we have assumed that each person changes at most one letter in the word. In general, we assume that there are *local conditions* that specify relations between values of the adjacent channels. In addition, for any $k \in \mathbb{Z}$, any $v \in V_k$, and any $p \in P_k$, we use predicate $Tr(v,p)$ to specify if an atomic proposition $p$ is "true" when the value of the channel $k$ is $v$.

DEFINITION 2. *A triple $(\{V_k\}_{k \in \mathbb{Z}}, \{L_k\}_{k \in \mathbb{Z}}, Tr)$ is called a protocol if*

1. $V_k$ *is an arbitrary set (of "values"), for each $k \in \mathbb{Z}$.*

2. $L_k \subseteq V_{k-1} \times V_k$ *is an arbitrary ("local condition") predicate, for each $k \in \mathbb{Z}$.*

3. $Tr$ *is a binary predicate such that $Tr \subseteq \bigcup_{k \in \mathbb{Z}}(V_k \times P_k)$.*

DEFINITION 3. *For any protocol $(\{V_k\}_{k \in \mathbb{Z}}, \{L_k\}_{k \in \mathbb{Z}}, Tr)$, a run is an arbitrary function $r(k)$ on $\mathbb{Z}$ such that $r(k) \in V_k$ and $(r(k-1), r(k)) \in L_k$ for each $k \in \mathbb{Z}$.*

Next is the core definition of this paper. It formally defines the semantics of the modality $\Box_k$.

DEFINITION 4. *For any given protocol*

$$\mathcal{P} = (\{V_k\}_{k \in \mathbb{Z}}, \{L_k\}_{k \in \mathbb{Z}}, Tr),$$

*we define relation $\Vdash$ between an arbitrary run $r$ of the protocol $\mathcal{P}$ and an arbitrary formula $\varphi \in \Phi(\mathbb{Z})$ as follows:*

1. $r \nVdash \bot$,
2. $r \Vdash p$ *if $Tr(r(k), p)$, where $p \in P_k$.*
3. $r \Vdash \varphi \to \psi$ *if $r \nVdash \varphi$ or $r \Vdash \psi$,*
4. $r \Vdash \Box_k \varphi$ *if $r' \Vdash \varphi$ for each $r'$ such that $r'(k) = r(k)$.*

Note that relation $r'(k) = r(k)$ between runs $r'$ and $r$ is an equivalence relation. Thus, the set of all runs of any given protocol acts as a set of possible worlds of an $S_5$ Kripke frame.

## 3. AXIOMS

Our logical system is an extension of the multi-modal version of $S_5$ by additional properties that deal with atomic propositions and topological structure of the communication chain. As will be shown in the next section, the traditional transitivity and $S_5$ axioms of the modal logic $S_5$ follow from a more general[2] Self-Awareness axiom below.

1. Distributivity: $\Box_k(\varphi \to \psi) \to (\Box_k \varphi \to \Box_k \psi)$,
2. Reflexivity: $\Box_k \varphi \to \varphi$,
3. Self-Awareness: $\varphi \to \Box_k \varphi$, where $\varphi \in \Phi(\{k\})$,
4. Gateway: $\Box_k \varphi \to \Box_n \varphi$, where $\varphi \in \Phi(A)$ and either $k < n \leq min(A)$ or $max(A) \leq n < k$,
5. Disjunction: $\Box_k(\varphi \vee \psi) \to \Box_k \varphi \vee \Box_k \psi$, where $\varphi \in \Phi(A)$, $\psi \in \Phi(B)$, and $max(A) \leq k \leq min(B)$.

We write $\vdash \varphi$ if $\varphi \in \Phi(\mathbb{Z})$ is provable from the axioms above and propositional tautologies in the language $\Phi(\mathbb{Z})$ using the Modus Ponens inference rule and the Necessitation inference rule:

$$\frac{\varphi}{\Box_k \varphi}.$$

We write $X \vdash \varphi$ if $\varphi$ is provable from the theorems of our system and the additional set of axioms $X$ using only Modus Ponens inference rule.

---
[2]The Self-Awareness axiom includes, for example, the principle $p \to \Box_k p$ for $p \in P_k$, which is not provable in $S_5$.

## 4. EXAMPLES

Soundness of our logical system will be shown in the next section. Here we give several examples of proofs in our formal system.

PROPOSITION 1 (TRANSITIVITY). $\vdash \Box_k\varphi \to \Box_k\Box_k\varphi$ for each $\varphi \in \Phi(\mathbb{Z})$ and each $k \in \mathbb{Z}$.

PROOF. Note that $\Box_k\varphi \in \Phi(\{k\})$. Thus, by the Self-Awareness axiom, $\vdash \Box_k\varphi \to \Box_k\Box_k\varphi$. □

PROPOSITION 2 (S5 AXIOM). $\vdash \Diamond_k\varphi \to \Box_k\Diamond_k\varphi$, for each $\varphi \in \Phi(\mathbb{Z})$ and each $k \in \mathbb{Z}$.

PROOF. Note that $\Diamond_k\varphi \in \Phi(\{k\})$. Thus, by the Self-Awareness axiom, $\vdash \Diamond_k\varphi \to \Box_k\Diamond_k\varphi$. □

PROPOSITION 3. If $k \leq m \leq n$ and $\varphi \in \Phi(\mathbb{Z})$, then
$$\vdash \Box_k\Diamond_n\varphi \to \Box_m\Diamond_n\varphi.$$

PROOF. Note that $\Diamond_n\varphi \in \Phi(\{n\})$. Thus, by the Gateway axiom, $\vdash \Box_k\Diamond_n\varphi \to \Box_m\Diamond_n\varphi$. □

PROPOSITION 4. If $k \leq m \leq n$ and $\varphi \in \Phi(\mathbb{Z})$, then
$$\vdash \Box_k\Box_n\varphi \to \Box_k\Box_m\Box_n\varphi.$$

PROOF. Note that $\Box_n\varphi \in \Phi(\{n\})$. Hence, by the Gateway axiom, $\vdash \Box_k\Box_n\varphi \to \Box_m\Box_n\varphi$. Thus, by the Necessitation rule, $\vdash \Box_k(\Box_k\Box_n\varphi \to \Box_m\Box_n\varphi)$. Then, by the Distributivity axiom, $\vdash \Box_k\Box_k\Box_n\varphi \to \Box_k\Box_m\Box_n\varphi$. Therefore, $\vdash \Box_k\Box_n\varphi \to \Box_k\Box_m\Box_n\varphi$ by Proposition 1. □

PROPOSITION 5. If $k \leq m \leq n$ and $\varphi, \psi \in \Phi(\mathbb{Z})$, then
$$\vdash \Box_m(\Box_k\varphi \vee \Box_n\psi) \to (\Box_m\varphi \vee \Box_m\psi).$$

PROOF. Note that $\Box_k\varphi \in \Phi(\{k\})$ and $\Box_n\psi \in \Phi(\{n\})$. Hence, by the Disjunction axiom,
$$\vdash \Box_m(\Box_k\varphi \vee \Box_n\psi) \to (\Box_m\Box_k\varphi \vee \Box_m\Box_n\psi). \qquad (2)$$

At the same time, by the Reflexivity axiom, $\vdash \Box_k\varphi \to \varphi$. Hence, by the Necessitation rule, $\vdash \Box_m(\Box_k\varphi \to \varphi)$. Thus, by the Distributivity axiom, $\vdash \Box_m\Box_k\varphi \to \Box_m\varphi$. One can similarly show that $\vdash \Box_m\Box_n\psi \to \Box_m\psi$. Therefore, from Statement (2), $\vdash \Box_m(\Box_k\varphi \vee \Box_n\psi) \to (\Box_m\varphi \vee \Box_m\psi)$. □

## 5. SOUNDNESS

Soundness of propositional tautologies and the Modus Ponens inference rule is straightforward. We will prove soundness of the Necessitation rule and of the remaining five axioms as separate lemmas.

LEMMA 1 (NECESSITATION). If $r \Vdash \varphi$ for any run $r$ of any protocol, then $r \Vdash \Box_k\varphi$ for any run $r$ of any protocol.

PROOF. Consider any run $r$. It will be sufficient to show that $r' \Vdash \varphi$ for each $r'$ such that $r'(k) = r(k)$, which is true due to the assumption of the lemma. □

LEMMA 2 (DISTRIBUTIVITY). For any run $r$ of a protocol $P$, if $r \Vdash \Box_k(\varphi \to \psi)$ and $r \Vdash \Box_k\varphi$, then $r \Vdash \Box_k\psi$.

PROOF. Let $r'$ be any run of $P$ such that $r'(k) = r(k)$. We will show that $r' \Vdash \psi$. Indeed, by the first assumption, $r' \Vdash \varphi \to \psi$. By the second assumption, $r' \Vdash \varphi$. Therefore, by Definition 4, $r' \Vdash \psi$. □

LEMMA 3 (REFLEXIVITY). For any run $r$ of a protocol $P$, if $r \Vdash \Box_k\varphi$, then $r \Vdash \varphi$.

PROOF. Lemma follows from Definition 4 and the fact that $r(k) = r(k)$. □

In the proofs of the soundness of the next three axioms, we use the following auxiliary lemma:

LEMMA 4. For any $A \subseteq \mathbb{Z}$, any formula $\varphi \in \Phi(A)$, and any runs $r, r'$ of the protocol $(\{V_k\}_{k \in \mathbb{Z}}, \{L_k\}_{k \in \mathbb{Z}}, Tr)$ such that $r(a) = r'(a)$ for every $a \in A$, $r \Vdash \varphi$ if and only if $r' \Vdash \varphi$.

PROOF. Induction on structural complexity of formula $\varphi$. If $\varphi \equiv \bot$, then the required follows from Definition 4.

If $\varphi \equiv p \in P_a$ is an atomic proposition for some $a \in A$, then $r \Vdash p$, by Definition 4 is equivalent to $Tr(r(a), p)$. At the same time, $Tr(r(a), p)$ is equivalent to $Tr(r'(a), p)$ due to the assumption that $r(a) = r'(a)$. Finally, again by Definition 4, $Tr(r'(a), p)$ is equivalent to $r' \Vdash p$.

If $\varphi \equiv \varphi_1 \to \varphi_2$, then $r \Vdash \varphi_1 \to \varphi_2$ is equivalent to disjunction of $r \nVdash \varphi_1$ and $r \Vdash \varphi_2$ by Definition 4. The disjunction, by the Induction Hypothesis, is equivalent to the disjunction of $r' \nVdash \varphi_1$ and $r' \Vdash \varphi_2$. Which, in turn, is equivalent to $r' \Vdash \varphi_1 \to \varphi_2$ by Definition 4.

Finally, assume that $\varphi \equiv \Box_a\psi$ for some $a \in A$. Without loss of generality, we suppose $r \Vdash \Box_a\psi$ and will prove $r' \Vdash \Box_a\psi$. Indeed, let $r''$ be any run of the protocol such that $r''(a) = r'(a)$. It will be sufficient to show that $r'' \Vdash \psi$. Note that $r''(a) = r'(a) = r(a)$. Thus, $r'' \Vdash \psi$ due to the assumption $r \Vdash \Box_a\psi$ and Definition 4. □

LEMMA 5 (SELF-AWARENESS). For any run $r$ of a protocol $P$, any $k \in \mathbb{Z}$, and any $\varphi \in \Phi(\{k\})$, if $r \Vdash \varphi$, then $r \Vdash \Box_k(\varphi)$.

PROOF. Consider any run $r'$ such that $r'(k) = r(k)$. It will be sufficient to show that $r' \Vdash \varphi$, which is true due to the assumption $r \Vdash \varphi$ and Lemma 4. □

LEMMA 6 (GATEWAY). For any $A \subseteq \mathbb{Z}$, any $\varphi \in \Phi(A)$, any run $r$, and any $k, n \in \mathbb{Z}$ such that $k < n \leq min(A)$ or $max(A) \leq n < k$, if $r \Vdash \Box_k\varphi$, then $r \Vdash \Box_n\varphi$.

PROOF. Without loss of generality, assume that $k < n \leq min(A)$. Let $r'$ be any run such that $r(n) = r'(n)$. We will show that $r' \Vdash \varphi$. Indeed, consider function $r^+(x)$ on $\mathbb{Z}$ such that
$$r^+(x) = \begin{cases} r(x) & \text{if } x < n, \\ r'(x) & \text{otherwise.} \end{cases}$$

We will show that $r^+$ is a run of the protocol. It trivially satisfies local conditions $L_x$ for all $x \neq n$. To show that local condition $L_n$ is satisfied notice that $L_n(r^+(n-1), r^+(n))$ is equivalent to $L_n(r(n-1), r'(n))$. Then it is also equivalent to $L_n(r(n-1), r(n))$ due to the assumption $r(n) = r'(n)$. Statement $L_n(r(n-1), r(n))$ is true because $r$ is a run of the protocol.

Note that $r^+(k) = r(k)$ by the assumption $k < n$. Thus, $r^+ \Vdash \varphi$ by the assumption $r \Vdash \Box_k\varphi$. Hence, $r' \Vdash \varphi$ by Lemma 4 and due to the fact that $r^+(a) = r'(a)$ for each $a \in A$. □

LEMMA 7 (DISJUNCTION). For any $A, B \subseteq \mathbb{Z}$, any $\varphi \in \Phi(A)$, any $\psi \in \Phi(B)$, any run $r$, and any integer $k \in \mathbb{Z}$ such that $max(A) \leq k \leq min(B)$, if $r \Vdash \Box_k(\varphi \vee \psi)$, then $r \Vdash \Box_k\varphi \vee \Box_k\psi$.

PROOF. Suppose that $r \nVdash \Box_k\varphi \vee \Box_k\psi$. Thus, by Definition 4, $r \nVdash \Box_k\varphi$ and $r \nVdash \Box_k\psi$. Hence, by Definition 4, there are runs $r_1$ and $r_2$ where $r_1(k) = r(k) = r_2(k)$ such that $r_1 \nVdash \varphi$ and $r_2 \nVdash \psi$.

Consider function $r^+(x)$ on $\mathbb{Z}$ such that

$$r^+(x) = \begin{cases} r_1(x) & \text{if } x \leq k, \\ r_2(x) & \text{if } x \geq k. \end{cases}$$

This function is well defined since $r_1(k) = r_2(k)$. It satisfies local conditions of the protocol since runs $r_1$ and $r_2$ do. Thus, $r^+$ is a run of the protocol. Note that $r^+(a) = r_1(a)$ for each $a \in A$ and $r^+(b) = r_1(b)$ for each $b \in B$. Hence, by Lemma 4, $r^+ \nVdash \varphi$ and $r^+ \nVdash \psi$. Thus, by Definition 4, $r^+ \nVdash \varphi \vee \psi$. This is a contradiction with the assumption $r \Vdash \Box_k(\varphi \vee \psi)$ and the fact that $r^+(k) = r_1(k) = r(k)$. □

## 6. COMPLETENESS

In this section we will prove the completeness of our logical system with respect to the semantics defined above. We start with two technical lemmas.

LEMMA 8. $\vdash \Box_k(\varphi \wedge \psi) \to (\Box_k\varphi \wedge \Box_k\psi)$.

PROOF. It will be sufficient to prove that $\vdash \Box_k(\varphi \wedge \psi) \to \Box_k\varphi$. Note that $(\varphi \wedge \psi) \to \varphi$ is a propositional tautology. Thus, $\vdash \Box_k((\varphi \wedge \psi) \to \varphi)$ by the Necessitation rule. Hence, $\vdash \Box_k(\varphi \wedge \psi) \to \Box_k\varphi$, by the Distributivity axiom. □

LEMMA 9. For any disjoint subsets $A \subseteq \mathbb{Z}$, $B \subseteq \mathbb{Z}$, any family of formulas $\{\varphi_i\}_{i \in A \cup B}$, and any $k \in \mathbb{Z}$ such that $max(A) \leq k \leq min(B)$,

$$\vdash \Box_k\left(\bigvee_{i \in A \cup B} \varphi_i\right) \to \left(\Box_k\left(\bigvee_{i \in A} \varphi_i\right) \vee \Box_k\left(\bigvee_{i \in B} \varphi_i\right)\right).$$

PROOF. Note the the following formula is a propositional tautology in our language:

$$\bigvee_{i \in A \cup B} \varphi_i \to \left(\bigvee_{i \in A} \varphi_i \vee \bigvee_{i \in B} \varphi_i\right).$$

Hence, by the Necessitation Rule,

$$\vdash \Box_k\left(\bigvee_{i \in A \cup B} \varphi_i \to \left(\bigvee_{i \in A} \varphi_i \vee \bigvee_{i \in B} \varphi_i\right)\right).$$

Thus, by the Distributivity axiom,

$$\vdash \Box_k\left(\bigvee_{i \in A \cup B} \varphi_i\right) \to \Box_k\left(\bigvee_{i \in A} \varphi_i \vee \bigvee_{i \in B} \varphi_i\right).$$

Therefore,

$$\vdash \Box_k\left(\bigvee_{i \in A \cup B} \varphi_i\right) \to \Box_k\left(\bigvee_{i \in A} \varphi_i\right) \vee \Box_k\left(\bigvee_{i \in B} \varphi_i\right),$$

by the Disjunction axiom. □

THEOREM 1. If $\nvdash \varphi$, then there is a protocol $P$ and a run $r$ of the protocol $P$ such that $r \nVdash \varphi$.

PROOF. Assume that $\nvdash \varphi$. Let $X_0$ be a maximal and consistent subset of $\Phi(\mathbb{Z})$ containing $\neg\varphi$. Let $\mathbb{X}$ be the set of all maximal consistent subsets of $\Phi(\mathbb{Z})$.

DEFINITION 5. For any $X, Y \in \mathbb{X}$ let $X \sim_k Y$ mean that $\psi \in X$ if and only if $\psi \in Y$ for each $\psi \in \Phi(\{k\})$.

LEMMA 10. For any $X \in \mathbb{X}$ and any $\psi$ such that $\Box_k\psi \notin X$, there is $Y \in \mathbb{X}$ such that $Y \sim_k X$ and $\neg\psi \in Y$.

PROOF. We will first show that the following set is consistent: $\{\sigma \in \Phi(\{k\}) \mid \sigma \in X\} \cup \{\neg\psi\}$. Indeed, let there be $\sigma_1, \ldots, \sigma_n \in \Phi(\{k\}) \cap X$ such that

$$\vdash \sigma_1 \to (\sigma_2 \to \cdots \to (\sigma_n \to \psi) \ldots).$$

By the Necessitation rule,

$$\vdash \Box_k(\sigma_1 \to (\sigma_2 \to \cdots \to (\sigma_n \to \psi) \ldots)).$$

By multiple applications of the Distributivity axiom,

$$\vdash \Box_k\sigma_1 \to (\Box_k\sigma_2 \to \cdots \to (\Box_k\sigma_n \to \Box_k\psi) \ldots).$$

By multiple applications of the Self-Awareness axiom,

$$\vdash \sigma_1 \to (\sigma_2 \to \cdots \to (\sigma_n \to \Box_k\psi) \ldots).$$

Hence, by multiple applications of the Modus Ponens rule, $\sigma_1, \sigma_2, \ldots, \sigma_n \vdash \Box_k\psi$. Thus, $X \vdash \Box_k\psi$, which is a contradiction with maximality of $X$ and the assumption $\Box_k\psi \notin X$. Let $Y$ be a maximal consistent set containing $\{\sigma \in \Phi(\{k\}) \mid \sigma \in X\} \cup \{\neg\psi\}$.

We are only left to show that if $\sigma \in Y$, then $\sigma \in X$ for each $\sigma \in \Phi(\{k\})$. Indeed, assume that $\sigma \notin X$. Then, $\neg\sigma \in X$ by the maximality of $X$. Hence, $\neg\sigma \in Y$ due to the choice of $Y$. Therefore, $\sigma \notin Y$ due to consistency of $Y$. □

LEMMA 11. $\sim_k$ is an equivalence relation on $\mathbb{X}$, for each $k \in \mathbb{Z}$. □

We now will define protocol $P = (\{V_k\}_{k \in \mathbb{Z}}, \{L_k\}_{k \in \mathbb{Z}}, Tr)$.

DEFINITION 6. Let $V_k$ be the set of equivalence classes of $\mathbb{X}$ with respect to relation $\sim_k$.

By $[X]_k$ we mean the equivalence class of element $X$ with respect to the equivalence relation $\sim_k$.

DEFINITION 7. $L_k(\alpha, \beta)$ if set $\alpha \cap \beta$ is not empty.

DEFINITION 8. For any $p \in P_k$ and any set $X \in \mathbb{X}$, $Tr([X]_k, p)$ is true if $p \in Y$ for each $Y \sim_k X$.

In other words, $Tr([X]_k, p)$ iff $p \in \cap [X]_k$. This concludes the definition of the protocol $P$.

LEMMA 12. For each $\psi \in \Phi(A)$, any run $r$ of the protocol $P$, any $k \in \mathbb{Z}$, and any $X \in \mathbb{X}$, if $\Box_k\psi \in X$, $X \in r(k)$, and either $k \leq n \leq min(A)$ or $max(A) \leq n \leq k$, then $\Box_n\psi \in Z$ for each $Z \in r(n)$.

PROOF. Without loss of generality, let $k \leq n \leq min(A)$. Induction on $n$. If $n = k$, then $\Box_k\psi \in X$ implies, by Definition 5, that $\Box_k\psi \in Z$ for each $Z \sim_k X$. Therefore, $\Box_k\psi \in Z$ for each $Z \in r(k)$.

Assume now that $k < n$. By $L_n$ condition, there exists $Y$ such that $Y \in r(n-1) \cap r(n)$. By the Induction Hypothesis, $\Box_{n-1}\psi \in Y$. Hence, by the Gateway axiom, $Y \vdash \Box_n\psi$. Hence, $\Box_n\psi \in Y$ by maximality of $Y$. Thus, by the Definition 5, $\Box_n\psi \in Z$ for each $Z \sim_n Y$. Therefore, $\Box_n\psi \in Z$ for each $Z \in r(n)$. □

Recall that value of any run $r$ under protocol $P$ is an equivalence class of $\mathbb{X}$. Thus, $\cap\, r(k)$ is a set of formulas. We will refer to this intersection in the next lemma.

LEMMA 13. *For any non-empty set $A \subseteq \mathbb{Z}$ and any set of formulas $\{\psi_a\}_{a \in A}$ such that $\psi_a \in \Phi(\{a\})$ for each $a \in A$ and any $X \in r(k)$, if*
$$\Box_k \bigvee_{a \in A} \psi_a \in X$$
*and either $k \leq min(A)$ or $max(A) \leq k$, then there is $a_0 \in A$ such that $\psi_{a_0} \in \cap\, r(a_0)$.*

PROOF. Without loss of generality, assume $k \leq min(A)$. We will prove the lemma by induction on the size of set $A$.

Base Case. Suppose that $A = \{a_0\}$. By Lemma 12, assuming $n = a_0$, we have $\Box_{a_0} \psi_{a_0} \in X$ for each $X \in r(a_0)$. Hence, due to maximality of the set $X$ and the Reflexivity axiom, $\psi_{a_0} \in X$ for each $X \in r(a_0)$. Therefore, $\psi_{a_0} \in \cap\, r(a_0)$.

Induction Step. Suppose that $|A| > 1$. Let $X_0$ be any set from $r(min(A))$. By Lemma 12, assuming $n = min(A)$, we have
$$\Box_{min(A)} \bigvee_{a \in A} \psi_a \in X_0.$$
Hence, by Lemma 9 and due to maximality of $X_0$,
$$\Box_{min(A)} \left( \psi_{min(A)} \vee \bigvee_{a \in A \setminus \{min(A)\}} \psi_a \right) \in X_0.$$
By the Disjunction axiom,
$$X_0 \vdash \Box_{min(A)} \psi_{min(A)} \vee \Box_{min(A)} \bigvee_{a \in A \setminus \{min(A)\}} \psi_a.$$
Hence, due to maximality of the set $X_0$, one of the following statements is true:
$$\Box_{min(A)} \psi_{min(A)} \in X_0,$$
$$\Box_{min(A)} \bigvee_{a \in A \setminus \{min(A)\}} \psi_a \in X_0.$$
In either case, the required follows from the Induction Hypothesis. □

LEMMA 14. *For any non-empty set $A \subseteq \mathbb{Z}$ and any set of formulas $\{\psi_a\}_{a \in A}$ such that $\psi_a \in \Phi(\{a\})$ for each $a \in A$ and any $X \in r(k)$, if*
$$\Box_k \bigvee_{a \in A} \psi_a \in X,$$
*then there is $a_0 \in A$ such that $\psi_{a_0} \in \cap\, r(a_0)$.*

PROOF. By Lemma 9 and due to maximality of $X$,
$$\Box_k \left( \bigvee_{\substack{a \in A \\ a \leq k}} \psi_a \vee \bigvee_{\substack{a \in A \\ a > k}} \psi_a \right) \in X.$$
By the Disjunction axiom,
$$X \vdash \Box_k \left( \bigvee_{\substack{a \in A \\ a \leq k}} \psi_a \right) \vee \Box_k \left( \bigvee_{\substack{a \in A \\ a > k}} \psi_a \right).$$
Hence, due to maximality of the set $X$, one of the following statements is true:
$$\Box_k \left( \bigvee_{\substack{a \in A \\ a \leq k}} \psi_a \right) \in X \quad \text{or} \quad \Box_k \left( \bigvee_{\substack{a \in A \\ a > k}} \psi_a \right) \in X.$$
In either case, the required follows from Lemma 13. □

LEMMA 15. *$r \Vdash \psi$ if and only if $\psi \in \cap\, r(k)$, for each $k \in \mathbb{Z}$, each run $r$ of the protocol $P$, and each $\psi \in \Phi(\{k\})$.*

PROOF. Induction on structural complexity of formula $\psi$. If $\psi \equiv \bot$, then the required follows from consistency of the set $r(k)$ and Definition 4. If $\psi$ is an atomic proposition, then the required follows from Definition 8.

Assume that $\psi \equiv \sigma \to \sigma'$ for some $\sigma, \sigma' \in \Phi(\{k\})$.
($\Rightarrow$) : Suppose that $r \Vdash \sigma \to \sigma'$. Thus, $r \nVdash \sigma$ or $r \Vdash \sigma'$. In the first case, by the Induction Hypothesis, $\sigma \notin \cap\, r(k)$. Hence, there is $X \in r(k)$ such that $\sigma \notin X$. Thus, $\sigma \to \sigma' \in X$ due to maximality of the set $X$. Hence, by Definition 5, $\sigma \to \sigma' \in Y$, for each $Y \sim_k X$. Therefore, $\sigma \to \sigma' \in \cap\, r(k)$.

In the second case, by the Induction Hypothesis, $\sigma' \in \cap\, r(k)$. Thus, $\sigma' \in X$ for each $X \in r(k)$. Hence, $\sigma \to \sigma' \in X$ for each $X \in r(k)$ due to maximality of set $X$. Therefore, $\sigma \to \sigma' \in \cap\, r(k)$.

($\Leftarrow$) : Suppose that $r \nVdash \sigma \to \sigma'$. Thus, $r \Vdash \sigma$ and $r \nVdash \sigma'$. Then, by the Induction Hypothesis, $\sigma \in \cap\, r(k)$ and $\sigma' \notin \cap\, r(k)$. Hence, there is $X \in r(k)$ such that $\sigma \in X$ and $\sigma' \notin X$. Thus, by maximality of the set $X$ and the Modus Ponens rule, $\sigma \to \sigma' \notin X$. Therefore, $\sigma \to \sigma' \notin \cap\, r(k)$.

Finally, assume that $\psi_k \equiv \Box_k \sigma$. Let $\bigwedge_i \bigvee_j \sigma_j^i$ be the conjunctive normal form of the formula $\sigma$ such that $\sigma_j^i \in \Phi(\{j\})$ for each $i$ and each $j$. Note that the following formula is provable in propositional logic without any additional modal axioms:
$$\sigma \to \bigwedge_i \bigvee_j \sigma_j^i.$$
Thus, by the Necessitation Rule,
$$\vdash \Box_k \left( \sigma \to \bigwedge_i \bigvee_j \sigma_j^i \right).$$
By the Distributivity axiom,
$$\vdash \Box_k \sigma \to \Box_k \left( \bigwedge_i \bigvee_j \sigma_j^i \right). \tag{3}$$
One can similarly show that
$$\vdash \Box_k \left( \bigwedge_i \bigvee_j \sigma_j^i \right) \to \Box_k \sigma. \tag{4}$$
($\Leftarrow$) : Suppose that $\Box_k \sigma \in \cap\, r(k)$. Let $r'$ be any run of the protocol such that $r(k) = r'(k)$. We will show that $r' \Vdash \bigwedge_i \bigvee_j \sigma_j^i$.

Note that $\Box_k \sigma \in \cap\, r(k)$ implies that $\Box_k \sigma \in \cap\, r'(k)$, because of the assumption $r(k) = r'(k)$. Hence, $\Box_k \sigma \in X$ for each $X \in r'(k)$. Thus, taking into account Statement (3),
$$\Box_k \left( \bigwedge_i \bigvee_j \sigma_j^i \right) \in X.$$

Then, by Lemma 8,
$$\Box_k \left( \bigvee_j \sigma_j^i \right) \in X.$$
for each $X \in r'(k)$ and each $i$. Hence by Lemma 14, for each $i$ there is $j_0$ such that $\sigma_{j_0}^i \in \cap\, r'(j_0)$. Thus, by the Induction Hypothesis, for each $X \in r'(k)$ and each $i$ there is $j_0$ such that $r' \Vdash \sigma_{j_0}^i$. Hence, $r' \Vdash \bigwedge_i \bigvee_j \sigma_j^i$.

$(\Rightarrow)$ : Suppose that $\Box_k \sigma \notin \cap\, r(k)$. Thus, there is $X \in r(k)$ such that $\Box_k \sigma \notin X$. Then, due to Statement (4),
$$\Box_k \left( \bigwedge_i \bigvee_j \sigma_j^i \right) \notin X.$$
Hence, by Lemma 10, there is $Y \sim_k X$ such that
$$\neg \bigwedge_i \bigvee_j \sigma_j^i \in Y.$$
Thus, due to the maximality of $Y$, there is $i_0$ such that
$$\neg \bigvee_j \sigma_j^{i_0} \in Y.$$
Hence, due to the maximality of $Y$, for each $j$,
$$\neg \sigma_j^{i_0} \in Y. \qquad (5)$$
Consider function $r_Y$ such that $r_Y(n) = [Y]_n$ for each $n \in \mathbb{Z}$. Note that $Y \in [Y]_{n-1} \cap [Y]_n$. Thus, $[Y]_{n-1} \cap [Y]_n$ is not empty for any $n \in \mathbb{Z}$. Therefore, $r$ is a run of the protocol $P$. By Definition 5, Statement (5) implies that $\neg \sigma_j^{i_0} \in Y'$ for each $j$ and each $Y' \sim_j Y$. Hence, $\neg \sigma_j^{i_0} \in \cap\, r_Y(j)$ for each $j$. Thus, by the Induction Hypothesis, $r_Y \Vdash \neg \sigma_j^{i_0}$ for each $j$. Then,
$$r_Y \Vdash \neg \bigvee_j \sigma_j^{i_0}.$$
Hence,
$$r_Y \Vdash \neg \bigwedge_i \bigvee_j \sigma_j^i.$$
In other words, $r_Y \Vdash \neg \sigma$. Therefore, $r \nVdash \Box_k \sigma$. $\square$

To finish the proof of the theorem, assume that $\bigwedge_i \bigvee_j \varphi_j^i$ is the conjunctive normal form of the formula $\neg \varphi$ such that $\varphi_j^i \in \Phi(\{j\})$ for each $i$ and each $j$. Consider $r$ such that $r(n) = [X_0]_n$ for each $n \in \mathbb{Z}$. Note that $X_0 \in [X_0]_{n-1} \cap [X_0]_n$. Thus, $[X_0]_{n-1} \cap [X_0]_n$ is not empty for any $n \in \mathbb{Z}$. Therefore $r$ is a run of the protocol $P$.

Recall that $\neg \varphi \in X_0$. Thus, $\bigwedge_i \bigvee_j \varphi_j^i \in X_0$. Hence, $\bigvee_j \varphi_j^i \in X_0$ for each $i$ due to maximality of the set $X_0$. Hence, again due to maximality of $X_0$, for each $i$ there is $j_i$ such that $\varphi_{j_i}^i \in X_0$. Hence, by Lemma 15, $r \Vdash \varphi_{j_i}^i$ for each $i$. Thus, $r \Vdash \bigwedge_i \bigvee_j \varphi_j^i$. Therefore, $r \Vdash \neg \varphi$. In other words, $r \nVdash \varphi$. $\square$

## 7. CONCLUSIONS

### 7.1 Directed Chains

Although edges representing channels $a$, $b$, and $c$ in Figure 1 are drawn as directed, none of our definitions so far have used them as such. The "directness" of these edges can be captured by restricting the class of all protocols to these that satisfy the additional *continuity condition* [1]: for each $v \in V_{k-1}$ there is $u \in V_k$ such that $L_k(v, u)$. This requirement, however, does not change any of our results and the existing proof of completeness still holds because the protocol constructed in the proof of completeness satisfies the continuity condition. Indeed, for any $[X]_{k-1} \in V_{k-1}$ one can just take $[X]_k \in V_k$ and notice that $L_k$ is true because $X \in [X]_{k-1} \cap [X]_k$.

### 7.2 Communication Networks

Communication chains can be generalized to non-linear communication networks like the one depicted in Figure 3. Intuitively it is clear that if $\Box_a \Box_f \varphi$ on this network, then

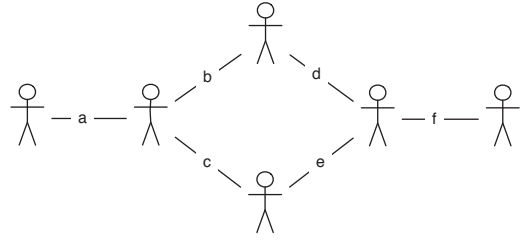

**Figure 3: Communication Network.**

this knowledge of $a$ is acquired through channels $b$ and $c$. This is an example of a more general form of the Gateway axiom for communication networks. However, straightforward formalization of this principle
$$\Box_a \Box_f \varphi \to (\Box_b \Box_f \varphi \vee \Box_c \Box_f \varphi)$$
is not true since the encrypted evidence of $\Box_f \varphi$ could have traveled through channels $b$ and $d$ and the encryption key through channels $c$ and $e$. Thus, neither $b$ nor $c$ alone would have knowledge of $\Box_f \varphi$ under such a protocol. The right way to formalize the Gateway principle in this setting is
$$\Box_a \Box_f \varphi \to (\Box_{b,c} \Box_f \varphi),$$
where $\Box_{b,c} \psi$ means that anyone who knows values $b$ and $c$ will be able to conclude $\psi$. In general, Definition 4 can be modified to say

4. $r \Vdash \Box_A \varphi$ if $r' \Vdash \varphi$ for each $r'$ such that $r'(a) = r(a)$ for all $a \in A$.

Then, the Gateway axiom can be stated as follows: if every path from each edge in set $A$ to each edge in set $B$ goes through an edge in set $G$, then
$$\Box_A \varphi \to \Box_G \varphi,$$
for each $\varphi \in \Phi(B)$. Similarly, the Disjunction axiom can be rephrased for communication networks as: if every path from each edge in set $A$ to each edge in set $B$ goes through an edge in set $G$, then
$$\Box_G (\varphi \vee \psi) \to (\Box_G \varphi) \vee (\Box_G \psi),$$
for each $\varphi \in \Phi(A)$ and each $\psi \in \Phi(B)$.

Both of these axioms are sound in the stated form. However, our proof of completeness heavily relies on equivalence relation $\sim_k$ and it is not clear how relations $\sim_A$ for multiple $A$ all of which might contain $k$ should work together. Thus, a complete axiomatization of epistemic logic for non-linear communication networks remains an open problem.